\documentclass[letterpaper,journal]{IEEEtran}
\usepackage{amssymb}
\usepackage[cmex10]{amsmath}
\usepackage{stfloats}
\usepackage{graphicx}
\usepackage{subfigure}
\usepackage{tabularx}
\usepackage{epsfig,epsf,color,balance,cite}
\usepackage{verbatim}
\usepackage{mathrsfs}
\usepackage{bm}
\usepackage{framed}

\usepackage{algorithm}
\usepackage{algpseudocode}
\usepackage{amsmath}
\usepackage{threeparttable}
\usepackage{diagbox}
\usepackage{booktabs,graphicx}
\usepackage{epstopdf}
\usepackage{epsfig}
\usepackage{booktabs}
\usepackage{array} 
\usepackage{mdframed}

\usepackage[table]{xcolor}
\definecolor{darkgreen}{rgb}{0.15 0.9 0.15} 

\begin{document}
\title{{Generative Semantic Communication: Architectures, Technologies, and Applications}}
\author{Jinke Ren, Yaping Sun, Hongyang Du, Weiwen Yuan, Chongjie Wang, Xianda Wang, \\Yingbin Zhou, Ziwei Zhu, Fangxin Wang, and Shuguang Cui,~\IEEEmembership{Fellow,~IEEE}
\thanks{J. Ren, W. Yuan, C. Wang, X. Wang, Y. Zhou, and Z. Zhu are with the Shenzhen Future Network of Intelligence Institute (FNii-Shenzhen), the School of Science and Engineering (SSE), and the Guangdong Provincial Key Laboratory of Future Networks of Intelligence, The Chinese University of Hong Kong, Shenzhen 518172, China (e-mail: jinkeren@cuhk.edu.cn; \{weiwenyuan1; chongjiewang; xiandawang; yingbinzhou; ziweizhu\}@link.cuhk.edu.cn).}
\thanks{Y. Sun is with the Department of Broadband Communication, Pengcheng Laboratory, Shenzhen 518000, China, and also with the FNii-Shenzhen, The Chinese University of Hong Kong, Shenzhen 518172, China (e-mail: sunyp@pcl.ac.cn).}
\thanks{H. Du is with the Department of Electrical and Electronic Engineering, The University of Hong Kong, Hong Kong, China (e-mail: duhy@eee.hku.hk). }
\thanks{F. Wang and S. Cui are with the SSE, the FNii-Shenzhen, and the Guangdong Provincial Key Laboratory of Future Networks of Intelligence, The Chinese University of Hong Kong, Shenzhen 518172, China (e-mail: wangfangxin@cuhk.edu.cn; shuguangcui@cuhk.edu.cn).} 
}
\pagestyle{headings} 
\maketitle 
\thispagestyle{empty}

\begin{abstract}
Semantic communication (SemCom) has emerged as a transformative paradigm for future wireless networks, promising to enhance communication efficiency by transmitting only the semantic meaning (or its coded version) of the source data rather than the entirety of bits (symbols). However, traditional deep learning-based SemCom systems suffer from limited generalization, low robustness, and insufficient reasoning capabilities due to the intrinsic discriminative nature of deep neural networks. To address these issues, generative artificial intelligence (GAI) is envisioned as a promising solution, offering significant advantages in learning complex data distributions, transforming data between high- and low-dimensional spaces, and creating high-quality content. 

This paper delves into the applications of GAI in SemCom and presents a thorough study. Three popular SemCom systems enabled by classical GAI models are first introduced, including variational autoencoders, generative adversarial networks, and diffusion models. For each system, the fundamental concept of the GAI model, the corresponding SemCom architecture, and the associated literature review of recent efforts are elucidated. Then, a novel generative SemCom system is proposed by incorporating the cutting-edge GAI technology---large language models (LLMs). This system features two LLM-based AI agents at both the transmitter and receiver, serving as ``brains” to enable powerful information understanding and content regeneration capabilities, respectively. This innovative design allows the receiver to directly generate the desired content, instead of recovering the bit stream, based on the coded semantic information conveyed by the transmitter. Therefore, it shifts the communication mindset from ``information recovery" to ``information regeneration" and thus ushers in a new era of generative SemCom. A case study on point-to-point video retrieval is presented to demonstrate the superiority of the proposed generative SemCom system, showcasing a 99.98\% reduction in communication overhead and a 53\% improvement in retrieval accuracy compared to the traditional communication system. Furthermore, four typical application scenarios for generative SemCom are delineated, followed by a discussion of three open issues warranting future investigation. In a nutshell, this paper provides a holistic set of guidelines for applying GAI in SemCom, paving the way for the efficient implementation of generative SemCom in future wireless networks.
\end{abstract}

\section{Introduction}
The rapid advancements in artificial intelligence (AI) and big data technologies are driving unprecedented transformations in communication networks. Unlike the previous five generations of wireless networks, which were designed with a focus on rate-centric metrics such as throughput, latency, spectral efficiency, and connection density, the sixth generation (6G) wireless networks will integrate advanced information processing technologies to reshape global information transmission, ushering in a new era of interconnectedness among humans, machines, and environments \cite{saad2019vision}. However, the classical Shannon-based communication paradigm emphasizes the accurate transmission of data at the bit level, ensuring that the receiver can precisely reproduce the original data regardless of its underlying meaning and utility \cite{shannon1948mathematical}. Such a separation principle of data communication from data application inevitably leads to the transmission of redundant data, inherently carrying inefficiencies in resource utilization. Consequently, it struggles to meet the stringent transmission requirements of future intelligent applications, such as metaverse, autonomous driving, smart cities, and holographic communication.

Semantic communication (SemCom) has emerged as a promising technology to revolutionize the communication paradigm. The concept of SemCom was first proposed by Warren Weaver in 1953, which targets effective information transmission at the semantic level, rather than at the bit level, thus enhancing communication efficiency \cite{weaver1953recent}. Recently, the academic community has materialized this concept by developing dedicated designs for specific tasks, also known as task-oriented communication \cite{gunduz2022beyond,wen2023task}. The working mechanism of SemCom typically involves four stages: initially, the receiver sends task requirements to the transmitter via a backward channel; subsequently, the transmitter extracts task-relevant semantic information from the source data and encodes it into signals suited to the channel condition; following this, the transmitter sends the coded signals to the receiver through a forward channel; finally, the receiver decodes the received noisy signals and reproduces semantically equivalent information. By filtering out task-irrelevant information before transmission, SemCom is able to significantly reduce the communication overhead \cite{yang2022semantic}.

Due to the groundbreaking progress in deep learning (DL) technologies, recent years have witnessed a proliferation of research in DL-based SemCom, covering key topics such as text transmission \cite{xie2021deep}, image reconstruction \cite{huang2022toward}, and speech recognition \cite{weng2023deep}. Despite these exciting achievements, the deployment of such systems in real-world scenarios faces three challenges: {\textit{1) Limited generalization capability:} existing works often employ hand-engineered neural networks for customized designs tailored to specific data modalities and tasks. These ``one-by-one" approaches require extensive human efforts for implementation and validation, and may not perform well in multimodal and multitask scenarios; {\textit{2) Low robustness:} current SemCom systems typically require end-to-end training under a particular network environment, rendering these systems incapable of rapid response and adaptive adjustment in dynamic environments with unseen data distributions; {\textit{3) Insufficient reasoning ability:} existing DL-based SemCom systems are restricted by the diversity of training data and lack sufficient background knowledge, making contextual reasoning and semantic calibration challenging at the receiver.

To overcome these challenges, generative artificial intelligence (GAI) is recognized as a potential technology that offers new opportunities for the development of SemCom. Broadly speaking, GAI encompasses intelligent technologies that generate original content---such as texts, images, and videos---by learning intrinsic patterns and latent features from a vast amount of data. Its exceptional capabilities arise from the deep integration of data and algorithms. On one hand, through self-supervised learning mechanisms, GAI can leverage large-scale unlabeled data to learn rich knowledge, thereby exhibiting strong generalization capabilities in cross-modal and multitask scenarios. On the other hand, GAI's robust network architecture can effectively capture rich semantics for information understanding and content generation, enabling adaptive solutions to complex tasks across different scenarios. Given the two advantages, there is a growing consensus that advocates for the application of GAI in SemCom \cite{yang2024streamlined, xia2023generative, liang2024generative, grassucci2024generativemag, ren2024knowledge}. Notably, our previous work \cite{ren2024knowledge} has introduced an innovative concept of ``generative SemCom" for the first time and constructed three GAI-empowered semantic knowledge bases (KBs) to enable efficient semantic coding and transmission.

The aforementioned studies demonstrate the potential of GAI to enhance the efficiency and flexibility of SemCom. To move one step further, this paper elucidates the intrinsic mechanisms by which GAI empowers SemCom. We first present a systematic review of SemCom research based on three classic GAI models, i.e., variational autoencoders (VAEs), generative adversarial networks (GANs), and diffusion models (DMs). Building upon this, we integrate the cutting-edge GAI technology---large language models (LLMs) and develop a new LLM-native generative SemCom system. This system offers a coherent design that harnesses the powerful capabilities of LLMs in information understanding and content generation to achieve the goal of SemCom. \textit{Its core lies in moving beyond the reconstruction-based ``encode-before-decode” mindset, adopting a new philosophy where the transmitter deeply understands the original data while the receiver directly generates the desired content based on the transmitter's understanding outcome.} This ``understand-before-regenerate" mindset is poised to reshape the fundamental architecture of future wireless networks and pave the last mile for the practical implementation of 6G.

The remainder of the paper is organized as follows. Section II introduces three types of SemCom systems enabled by VAEs, GANs, and DMs, respectively. Section III reviews the basic concept of LLMs and their existing applications in SemCom, followed by presenting the LLM-native generative SemCom system. Section IV provides a case study to demonstrate the effectiveness of the newly proposed generative SemCom system. Several application scenarios and research directions for generative SemCom are elaborated in Sections V and VI, respectively. Finally, Section VII concludes the paper.
\section{Traditional SemCom Empowered by GAI Models}
In this section, we will introduce three prominent SemCom systems, each built upon a distinct GAI model, i.e., VAEs, GANs, and DMs. 
\subsection{VAE-enabled SemCom}
\subsubsection{Introduction to VAEs}
VAEs are well-known probabilistic generative models introduced by Diederik P. Kingma and Max Welling in 2013~\cite{kingma2013auto}. The key idea of VAEs is to learn the probabilistic distribution of training data to generate new samples as variations of the input data. A standard VAE consists of a probabilistic encoder and a probabilistic decoder, typically implemented using neural networks. The encoder compresses input data into a low-dimensional latent space, parameterized by mean and standard deviation that characterizes the statistical properties of the input data. The decoder samples latent variables from this space to reconstruct the input data. A distinguishing feature of VAEs is their ability to quantify data uncertainty during training through a Kullback-Leibler divergence term. This term regularizes the latent space to make it conform to a standard normal distribution, hence establishing a continuous and smooth latent space for coherent sample generation.

The robust data generation capabilities of VAEs have led to their widespread use across various domains, such as image generation~\cite{razavi2019generating} and drug discovery~\cite{li2021co}. Notably, the ``compress-before-reconstruct" approach of VAEs aligns well with that of SemCom, enabling a significant reduction in communication overhead.
\begin{figure*}[t]
\centering
\includegraphics[width=0.73\linewidth]{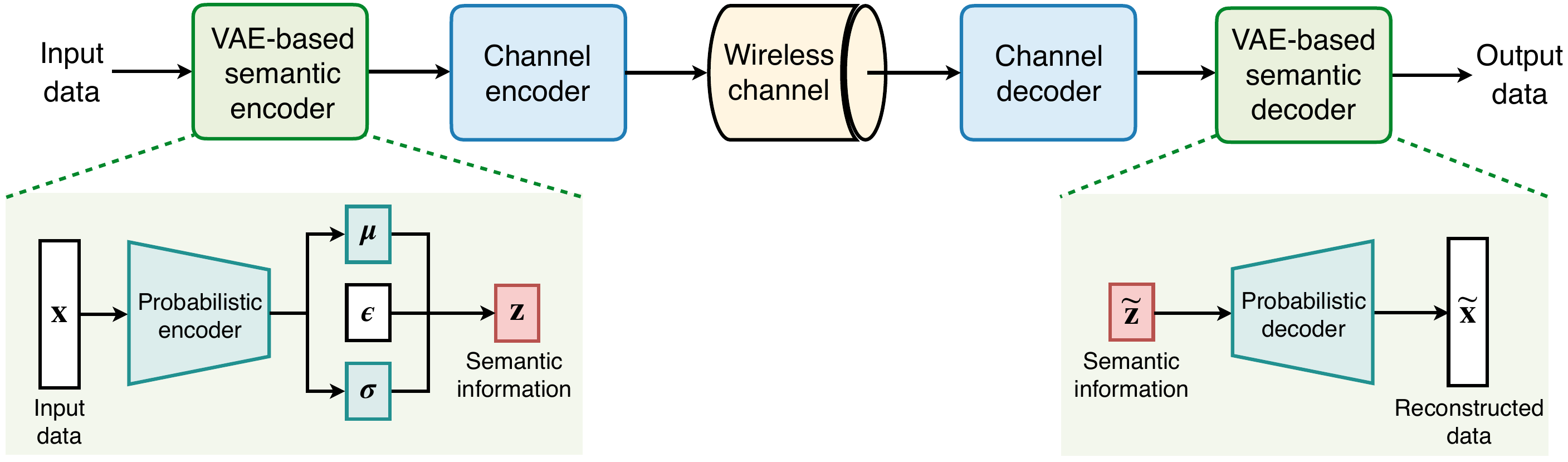}
\caption{VAE-enabled SemCom architecture.} \label{fig:VAE_architecture}
\vspace{-8pt}
\end{figure*}
\subsubsection{VAE-enabled SemCom Architecture}
The probabilistic coding scheme of VAEs promises efficient feature extraction and data reconstruction, inspiring a typical VAE-enabled SemCom architecture as shown in Fig. \ref{fig:VAE_architecture}. The probabilistic encoder and decoder of the VAE are respectively deployed at the transmitter and receiver, functioning as the semantic encoder and semantic decoder.
\begin{itemize}
    \item \textit{\textbf{Semantic encoding}:} At the transmitter, the VAE encoder performs semantic encoding by extracting essential statistical information from the input data. It first estimates the mean $\bm \mu$ and standard deviation $\bm{\sigma}$ of the latent space for the input data ${\mathbf{x}}$. Then, by utilizing the reparameterization trick, the transmitter samples a random vector $\bm \epsilon$ from a standard normal distribution $\mathcal{N}\left( 0, \mathbf{I}
    \right)$ and computes a low-dimensional latent vector by
    \begin{equation}
        \mathbf{z} = {\bm \mu} + {\bm \epsilon} \odot {\bm \sigma}^2,
    \end{equation}  
    where $\odot$ denotes element-wise multiplication. Consequently, the distribution of the latent vector approximates a normal distribution $\mathcal{N}({\bm \mu}, {\bm \sigma}^2)$. Notably, the latent vector $\mathbf{z}$ captures the statistical properties of the original data, 
    thereby encapsulating its semantic information. 
    \item \textit{\textbf{Semantic decoding}:} At the receiver, the VAE decoder performs semantic decoding to reconstruct the original data. It involves mapping the received noisy latent vector $\widetilde{\mathbf{z}}$ output by the channel decoder back into the high-dimensional data space, hence completing the communication task.
\end{itemize}
\subsubsection{Literature Review}
In recent years, VAEs and their variants have been widely employed in SemCom, particularly for semantic coding and joint source-channel coding (JSCC).
\paragraph{\textbf{\textit{Standard VAEs for coding}}} Following the introduced architecture, an early study \cite{saidutta2020vae} proposed a VAE-based JSCC scheme for distributed Gaussian sources over a multiple access Additive white Gaussian noise (AWGN) channel, which eliminated the necessity of the joint distribution of all sources in traditional JSCC schemes. A subsequent work \cite{erdemir2022privacy} developed a privacy-aware JSCC scheme for wireless wiretap channels, which leveraged the VAE encoder to map source data into a latent probability distribution rather than a deterministic representation, thereby confusing the eavesdropper's perception of privacy-sensitive attributes. Following it, Ref. \cite{alawad2022innovative} explored a short packet communication system, which transmitted only the statistical parameters of packets to enhance spectral efficiency. Additionally, Refs. \cite{xi2024variational} and \cite{yao2023variational} investigated two scenarios of text and speech transmission. Ref. \cite{xi2024variational} integrated variational neural inference and conventional explicit semantic decoding scheme to improve the text decoding accuracy. Ref. \cite{yao2023variational} employed nonlinear transformations and variational modeling to capture dependencies between speech frames and estimate the probabilistic distribution of speech features, thus optimizing the speech encoding and reconstruction processes. Furthermore, a VAE-based joint coding-modulation framework was proposed in Ref. \cite{bo2024joint}, which mapped source data to discrete constellation symbols in a probabilistic manner, thereby resolving the non-differentiable issues in digital modulation. In particular, several recent studies have leveraged deep reinforcement learning (DRL) \cite{seon2024deep} and semantic-aware codebooks \cite{zhang2024improving} to provide side information to further enhance the semantic coding performance.}
 
\paragraph{\textbf{\textit{VAE variants for coding}}} To meet diverse task requirements, several pioneering efforts have utilized variants of VAEs, including hierarchical VAEs (HAVEs) \cite{chen2024hierarchical,zhang2024learned}, conditional VAEs (CVAEs) \cite{li2024end,xie2024robust}, $\beta$-VAEs \cite{ma2023task}, and vector quantized VAEs (VQ-VAEs) \cite{nemati2023vq,hu2023robust,talli2024effective,choi2024semantics,si2024post} to develop advanced coding schemes. HVAEs learn a hierarchical data representation through multiple latent variables, thus enhancing the flexibility of semantic coding. For instance, Ref. \cite{chen2024hierarchical} employed a HAVE to map source data into low-level and high-level latent variables, which capture local detail features and global abstract features, respectively. These latent variables were flexibly combined to improve the accuracy of remote wireless control tasks. Additionally, Ref. \cite{zhang2024learned} utilized a HAVE to autoregressively learn multiple latent representations of images through a combination of bottom-up and top-down paths. The latent representations were then mapped to different numbers of channel symbols for transmission, thereby achieving favorable rate-distortion performance. CVAEs introduce external conditioning variables to enable fine-grained control during data generation. For example, Refs. \cite{li2024end} and \cite{xie2024robust} incorporated image categories and channel state information (CSI) into the semantic coding process, enhancing data reconstruction accuracy and system's robustness against channel noise. Besides, by introducing a hyperparameter $\beta$ in the semantic coding process, the learned latent variables can be decoupled into a series of task-related independent variables, thus improving the interpretability of SemCom systems \cite{ma2023task}. Additionally, to align with digital communication systems, many studies \cite{nemati2023vq,hu2023robust,choi2024semantics,talli2024effective,si2024post} have employed VQ-VAEs in SemCom. By learning discrete latent representations of source data and mapping them to codewords within a shared codebook at the transmitter and receiver, the communication overhead is significantly reduced since only the indices of the selected codewords need to be transmitted. Among them, VQ-VAEs were augmented through adversarial training methods \cite{hu2023robust}, DRL algorithms \cite{talli2024effective}, and fine-tuning techniques \cite{choi2024semantics,si2024post}, improving systems' robustness against channel impairments and enhancing their generalization to unseen data distributions.
\subsection{GAN-enabled SemCom}
\subsubsection{Introduction to GANs}
GANs have emerged as prominent models in GAI, capable of generating realistic data by learning the statistical properties of a large amount of real data. Originally conceptualized by Ian Goodfellow and his colleagues in 2014~\cite{goodfellow2014generative}, a standard GAN comprises two neural networks, namely a generator and a discriminator, which are trained concurrently in an adversarial manner. Specifically, the generator takes random noise as input and is trained to synthesize data that mislead the discriminator into classifying them as real instances. Conversely, the discriminator is trained to accurately distinguish between real data and synthesized samples output by the generator. This adversarial training process can be mathematically formulated as a two-player minimax game, wherein the generator aims to minimize the discriminator's ability to differentiate between real and synthesized data, while the discriminator strives to maximize its classification accuracy. This competitive dynamic continues until the system reaches a Nash equilibrium, at which point the generator can produce highly plausible data while the discriminator is deceived approximately half the time.

The efficacy of GANs can be attributed to their proficiency in modeling intricate data distributions without necessitating explicit likelihood functions or prior distributions. Moreover, the ``evolutionary arms race" between the generator and discriminator provides new methodologies for SemCom, enabling the receiver to reproduce data that faithfully reflects the characteristics of the source input.
\begin{figure*}[t]
\centering
\includegraphics[width=0.75\linewidth]{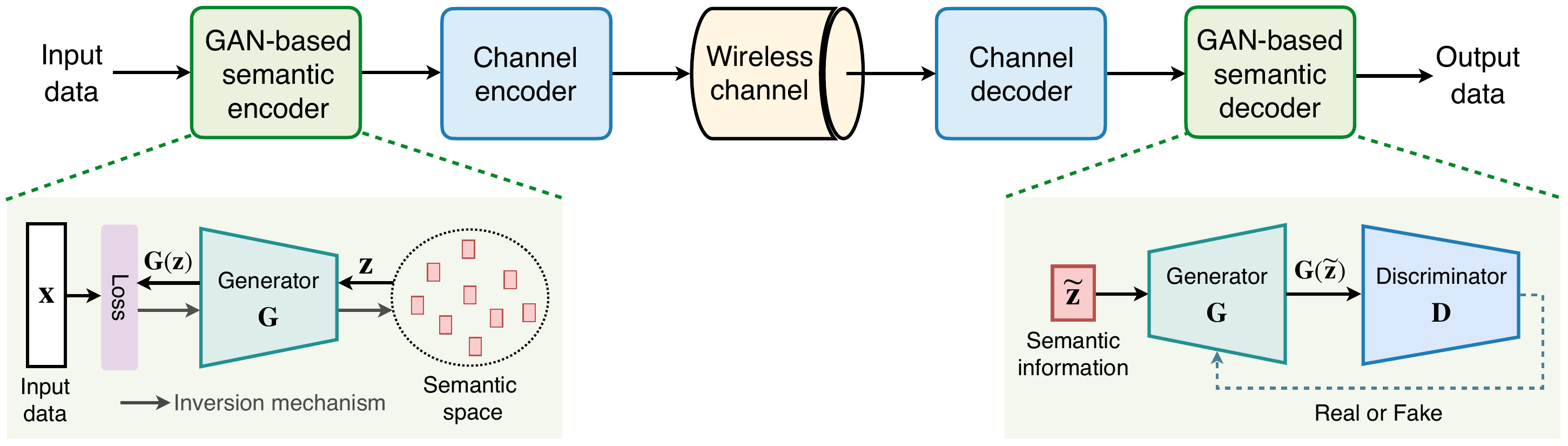}
\caption{GAN-enabled SemCom architecture.} \label{fig:GAN_architecture}
\vspace{-8pt}
\end{figure*}
\subsubsection{GAN-enabled SemCom Architecture}
Fig. \ref{fig:GAN_architecture} illustrates a typical GAN-enabled SemCom architecture, where GANs are employed for both semantic encoding at the transmitter and semantic decoding at the receiver.
\begin{itemize}
    \item \textit{\textbf{Semantic encoding}:} The generator of the GAN can act as the semantic encoder to extract semantic information from the input data through an inversion mechanism \cite{han2023generative}. Specifically, the transmitter samples a latent variable $\mathbf{z}$ from a predefined semantic space and inputs it into the generator to produce synthesized data $\mathbf{G}(\mathbf{z})$. The latent variable $\mathbf{z}$ is then iteratively optimized to minimize the discrepancy between the synthesized data $\mathbf{G}(\mathbf{z})$ and the input data $\mathbf{x}$. Oftentimes, the discrepancy is quantified using mean squared error (MSE) or learned perceptual image patch similarity (LPIPS). When using MSE, the optimization towards $\mathbf{z}$ can be expressed as
    \begin{equation}
       \mathbf{z}^\star = \arg \min_{\mathbf{z}} \left\Vert \mathbf{G}(\mathbf{z}) - \mathbf{x} \right\Vert_2^2,
    \end{equation}
    where $\left\Vert \cdot \right\Vert_2$ denotes the $L_2$ norm and $\mathbf{z}^\star$ is seen as the semantic information.
    \item \textit{\textbf{Semantic decoding}:} At the receiver, the generator of the GAN serves as the semantic decoder, which reconstructs data based on the received noisy semantic information, denoted by $\widetilde{\mathbf{z}}$. During the training phase, the reconstructed data $\mathbf{G}\left(\widetilde{\mathbf{z}}\right)$ is input into the discriminator for authenticity assessment. The result is subsequently utilized to update the generator, thereby improving the quality of the reconstructed data. This process is formalized as a min-max game given by
    \begin{equation}
    \begin{split}
        \min_{\mathbf{G}} \max_{\mathbf{D}}~V \left(\mathbf{D},\mathbf{G}\right) = \mathbb{E}_{\mathbf{x} \sim p_{\text{data}} \left( \mathbf{x}\right)} \left[ \log \mathbf{D}\left( \mathbf{x}\right) \right] \\ + \mathbb{E}_{ \widetilde{\mathbf{z}} \sim p_{\widetilde{\mathbf{z}}} \left(\widetilde{\mathbf{z}}\right)} \left[\log \left(1 - \mathbf{D}\left(\mathbf{G} \left( \widetilde{\mathbf{z}} \right) \right) \right)\right], 
    \end{split}
    \end{equation}
    where $p_{\text{data}} \left( \mathbf{x} \right)$ denotes the ground-truth data distribution and $p_{\widetilde{\mathbf{z}}} \left(\widetilde{\mathbf{z}}\right)$ is the distribution of the noisy semantic information. Through iterative optimization, the generator is able to achieve efficient data reconstruction in the inference phase.
\end{itemize}
\subsubsection{Literature Review}
Existing research on GAN-enabled SemCom can be categorized into two directions, i.e., leveraging GANs to enable the encoding process at the transmitter and the reconstruction process at the receiver.

\paragraph{\textbf{\textit{GANs for encoding}}} The application of GANs in the transmitter’s encoding process holds significant potential, particularly for images. For instance, an early study \cite{huang2021deep} introduced a coarse-to-fine semantic encoding approach by integrating the generator of GAN with the better portable graphics (BPG) residual encoding technology. This approach extracted multi-level semantic features from input images, enabling the receiver to reconstruct high-quality images with fine details. Following it, Refs. \cite{han2023generative} and \cite{tang2024evolving} employed the inversion mechanism of StyleGAN to extract semantic features from input images. Moreover, Ref. \cite{han2023generative} introduced a privacy filter to remove privacy-sensitive information from the semantic features, while Ref. \cite{tang2024evolving} established a semantic cache module to reduce redundant semantic transmission. Additionally, a CycleGAN-based data adaptation module was developed to align the input data with the pre-stored empirical data in the KB \cite{zhang2022deep}. By this means, the transmitter can perform adaptive semantic encoding without the need for retraining.

\paragraph{\textbf{\textit{GANs for reconstruction}}} The adversarial interaction between the generator and discriminator offers distinct advantages in the receiver’s decoding process \cite{huang2022toward}. Specifically, the generator serves as the semantic decoder or JSCC decoder to reconstruct data from the received noisy semantic information, whereas the discriminator engages in adversarial training with the generator to enhance reconstruction performance \cite{he2022robust,wang2022perceptual,xin2024deep,tan2024rate}. Building on this concept, Ref. \cite{wang2022perceptual} introduced both perceptual and adversarial losses to capture global semantic information and local texture information during training, while Refs. \cite{xin2024deep} and \cite{tan2024rate} explored the trade-off between signal distortion, perceptual quality, and transmission rate. Moreover, Ref. \cite{erdemir2023generative} proposed two StyleGAN-based JSCC schemes, namely InverseJSCC and GenerativeJSCC. The former modeled the traditional DeepJSCC process \cite{bourtsoulatze2019deep} as an unsupervised inverse problem and utilized the StyleGAN generator to solve it, thereby overcoming the distortion issue in DeepJSCC. The latter achieved high-quality image reconstruction by jointly training the StyleGAN-based JSCC encoder and decoder in an end-to-end manner. Besides, Ref. \cite{yu2024two} presented a two-way SemCom system, which leveraged weight reciprocity between transceivers to eliminate the need for information feedback during training. A conditional GAN was also utilized to estimate channel distributions therein. To further realize adaptive image reconstruction, several studies adopted semantic slicing \cite{dong2022semantic}, semantic segmentation \cite{lokumarambage2023wireless}, and VQ-based semantic codebooks \cite{miao2024semantic,zhou2024moc,fu2023vector} to perform flexible semantic encoding at the transmitter. Finally, a recent work \cite{mao2024gan} applied GANs to text transmission scenarios, wherein a GAN-based distortion suppression module was designed to mitigate the signal distortion issue due to the absence of CSI.
%
\begin{figure*}[t]
\centering
\includegraphics[width=0.72\linewidth]{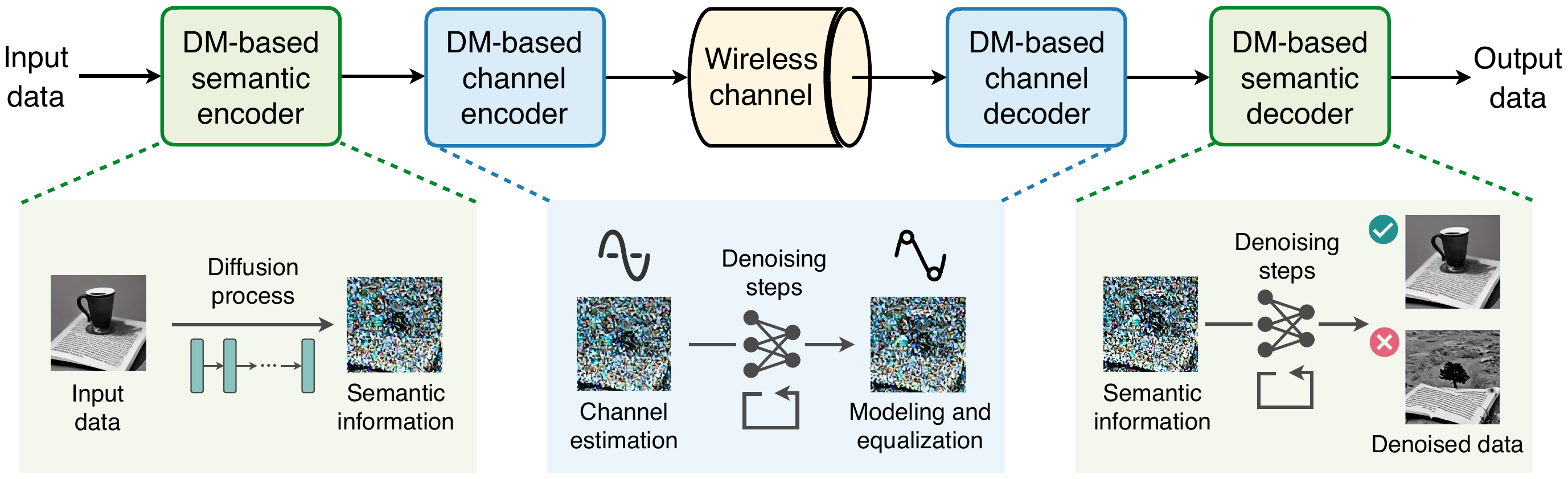}
\caption{DM-enabled SemCom architecture.} \label{fig:DM_architecture}
\vspace{-8pt}
\end{figure*}
\subsection{DM-enabled SemCom}
\subsubsection{Introduction to DMs}
DMs are a class of GAI models inspired by non-equilibrium thermodynamics, which have garnered significant attention for their abilities to model complex data distributions and generate high-fidelity samples~\cite{yang2023diffusion}. Typically, DMs encompass three types of formulations, i.e., denoising diffusion probabilistic models (DDPMs)~\cite{ho2020denoising}, score-based generative models (SGMs)~\cite{song2019generative}, and score-based stochastic differential equations (Score SDEs)~\cite{song2020score}. DDPMs operate through a two-step process involving a forward diffusion process and a reverse denoising process. The forward process progressively adds noise to the data to transform it into a noise distribution, while the reverse process learns to remove the noise from the noisy data to reconstruct the original data distribution. In contrast to the denoising method in DDPMs, SGMs generate new samples by estimating the score function of the noisy data distribution. Score SDEs further extend these methods to continuous time and model the data perturbation and generation process as solutions to well-formed SDEs. 

DMs excel in handling high-dimensional data and have achieved remarkable success in various applications, such as image generation \cite{rombach2022high} and time series forecasting \cite{kollovieh2024predict}. Notably, the inherent denoising ability of DMs makes them well-suited in eliminating the channel noise introduced during transmission in SemCom.
\subsubsection{DM-enabled SemCom Architecture}
Fig.~\ref{fig:DM_architecture} illustrates a typical DM-enabled SemCom architecture, where DMs empower the entire communication process from three aspects, including semantic encoding, channel modeling and equalization, and semantic decoding.
\begin{itemize}
    \item \textit{\textbf{Semantic encoding}:} At the transmitter, DMs are utilized to construct the semantic encoder, which produces a compact semantic representation of the input data through the diffusion process. This involves gradually adding Gaussian noise to the input data $\mathbf{x}$ over $T$ timesteps, following a forward Markov chain. At the timestep $t$, noise is added according to 
    \begin{equation}
        \mathbf{x}_t = \sqrt{\alpha_t}\mathbf{x}_{t-1} + \sqrt{1-\alpha_t}\boldsymbol{\epsilon}_{t-1},
    \end{equation}
    where $\mathbf{x}_0 = \mathbf{x}$, $\alpha_t$ is the noise schedule, and $\boldsymbol{\epsilon}_{t-1}$ is the noise sampled from a Gaussian distribution. Notably, the number of timesteps $T$ can be flexibly adjusted to strike a balance between data compression and semantic fidelity. More importantly, the DM-based semantic encoder exhibits strong robustness to input variations through the progressive noising process.
    \item \textit{\textbf{Channel modeling and equalization}:} The semantic information ${\bf x}_T$ is susceptible to various types of channel impairments (e.g., channel fading, interference, and noise) during wireless transmission. To address these issues, DMs can be utilized to facilitate channel modeling and equalization. Specifically, channel impairments can be seen as another form of noise addition to the semantic information. Therefore, DMs can perform denoising through a reverse Markov chain, which is expressed as  
    \begin{equation}
        \mathbf{\widetilde{x}}_{t-1} = \frac{1}{\sqrt{\alpha_t}}\left(\mathbf{x}_t - \frac{1-\alpha_t}{\sqrt{1-\bar{\alpha}_t}}\boldsymbol{\epsilon}_\theta(\mathbf{x}_t,t)\right),   
    \end{equation}
    where $\bar{\alpha}_t = \Pi_{i=1}^t \alpha_i$ and $\boldsymbol{\epsilon}_\theta (\cdot,\cdot)$ is a neural network for predicting the noise component.
    \item \textit{\textbf{Semantic decoding}:} At the receiver, DMs are harnessed to construct the semantic decoder, which reconstructs data via the reverse process. Through multiple denoising steps, the decoder gradually removes noise from received noisy signals and recovers the original data.
    DM-based semantic decoders work 
    well for generating high-dimensional data, such as images \cite{jiang2024diffsc,guo2024diffusion,chen2024commin,yuan2024generative} and audio \cite{qiang2024minimally}.
\end{itemize}
\subsubsection{Literature Review}
Recent studies on DM-enabled SemCom mainly focus on three key topics, including multimodal semantic coding, channel modeling and enhancement, as well as secure and efficient transmission.
\paragraph{\textbf{\textit{Multimodal semantic coding}}} For image transmission, an early study \cite{grassucci2023generative} deployed a DM at the receiver to generate semantically consistent images from received noisy semantic information. This concept was then extended to a multiuser scenario to overcome the issue of information loss caused by limited subcarrier availability \cite{grassucci2024rethinking}. Building upon the two studies, Ref. \cite{pignata2024lightweight} further applied post-training quantization to DMs, reducing their computational load and memory footprints for on-device deployment. Meanwhile, Ref. \cite{zhang2024sc} employed a compact DM to compute conditional vectors for enhancing decoding efficiency. On the other hand, some recent studies have employed semantic feature decomposition \cite{fan2024semantic, qiao2024latency} and RL techniques \cite{wang2024fast} to extract various types of image features (e.g., texture, color, and object), enabling the receiver-side DMs to perform flexible and controllable image reconstruction. Several studies have also incorporated semantic segmentation models to extract essential semantic maps at the transmitter \cite{pezone2024semantic,fu2024multimodal,mingkai2024task}. These maps served as guidance to facilitate DMs to generate high-quality images. Notably, deploying pre-trained semantic KBs at both the transmitter and receiver was demonstrated to further improve the quality of images generated by DMs \cite{jiang2024adaptive}. Moreover, DMs’ robust denoising and generation capabilities have led to their widespread application in practical SemCom scenarios, such as speech synthesis \cite{grassucci2024diffusion}, 3D object generation \cite{chen2024cross}, scene understanding \cite{yang2024sg2sc}, and panoramic image reconstruction in virtual reality \cite{zhang2024diffusion}.
\paragraph{\textbf{\textit{Channel modeling and enhancement}}} DM-based channel modeling and enhancement have demonstrated significant advantages in addressing channel impairments. Specifically, Ref. \cite{xu2023latent} first developed a latent diffusion-based denoising framework to eliminate noise interference during transmission. Following it, Ref. \cite{wu2024cddm} introduced a channel denoising DM as an add-on module to mitigate channel noise. Furthermore, Ref. \cite{li2024goal} proposed a stable diffusion-based denoiser that removed noise by learning the distribution of channel gains. Moreover, a consistency distillation strategy was developed in Ref. \cite{pei2024latent}, which transformed the multistep denoising process into a single or a few deterministic steps, thereby enabling real-time channel denoising. Besides, several recent studies have utilized DMs for CSI estimation and refinements, thereby enhancing the decoding performance at the receiver \cite{zeng2024dmce,jiang2024large}. Concurrently, a plug-in module named DM-MIMO was developed in \cite{duan2024dm}, which combined DMs and singular value decomposition techniques for precoding and channel equalization over multiple-input and multiple-output (MIMO) channels. In addition, rate-adaptive mechanisms \cite{yang2024rate} and hybrid digital-analog approaches \cite{xie2024hybrid} were also integrated with DMs to balance data rate and semantic distortion.
\paragraph{\textbf{\textit{Secure and efficient transmission}}} DMs hold substantial potential for improving the security and efficiency of SemCom systems. For instance, a DM-empowered secure scheme was proposed to overcome security risks posed by semantic-oriented attacks \cite{ren2024diffusion}. By adding Gaussian noise to original images and incorporating a denoising module to purify received images, this scheme significantly enhanced the system's robustness against adversarial perturbations. Ref. \cite{du2023ai} proposed a DM-based AI-generated contract mechanism to incentivize secure and efficient semantic sharing in a full-duplex device-to-device SemCom system. Besides, a DM-based defense mechanism was developed to resist potential adversarial attacks in semantic Internet-of-things (IoT) scenarios \cite{zheng2024energy}. In particular, this defense mechanism iteratively added and removed noise to negate adversarial perturbations in images, thereby achieving a balance between computational efficiency and network security. Furthermore, several studies have explored radio resource management to optimize the system-level efficiency of SemCom systems \cite{liu2024optimizing,du2023yolo,xu2024semantic}.
\section{LLM-driven Generative SemCom}
In this section, we explore SemCom systems enabled by the cutting-edge GAI models---LLMs. We first outline the key components of LLMs and present their existing applications in SemCom. Building upon this, we propose a novel LLM-native generative SemCom architecture to fully harness the generative potential of LLMs.
\subsection{Preliminaries of LLMs}
LLMs are a class of deep neural networks that are trained on a vast amount of data and possess a large scale of parameters. Compared to traditional AI models, LLMs not only possess comprehensive world knowledge but also exhibit exceptional understanding and reasoning capabilities, enabling them to execute complex tasks following human instructions and generate human-like responses \cite{zhao2023survey}. To date, LLMs have achieved astonishing success across a wide range of research domains and have become the engine for propelling technological advancement \cite{wan2023efficient}.

The predominant architectures of LLMs are categorized into three types: the encoder-only architecture exemplified by bidirectional encoder representations from transformers (BERT) \cite{devlin2018bert}, which excels in text understanding; the decoder-only architecture represented by generative pre-trained transformer (GPT) \cite{achiam2023gpt}, specializing in content generation; and the encoder-decoder architecture typified by text-to-text transfer transformer (T5) \cite{raffel2020exploring}, capable of handling both text understanding and content generation. All of these architectures are built on a foundational model called \textit{transformer block}. As depicted in Fig. \ref{fig:transformer_block}, a transformer block is composed of three components, including a multi-head self-attention (MHSA) mechanism, a feedforward neural network (FFN), and two layer normalizations with residual connections.
\begin{figure}[t]
\centering
\includegraphics[width=0.95\linewidth]{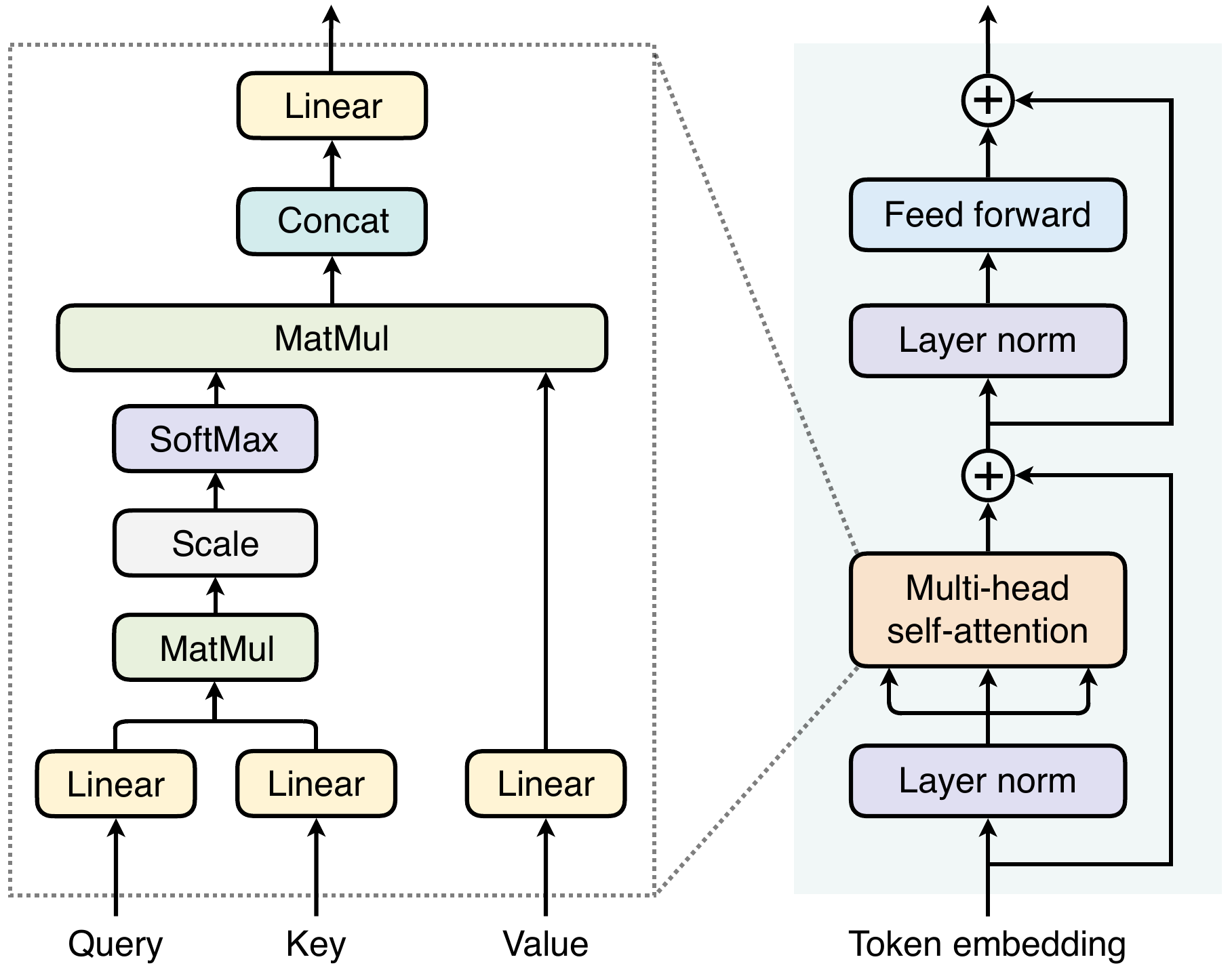}
\caption{An illustration of the transformer block. The left dotted box shows the components of a single attention layer, including linear projection (Linear), matrix multiplication (MatMul), scaling operation (Scale), softmax function (SoftMax), and concatenation (Concat).} 
\label{fig:transformer_block}
\end{figure}
\begin{itemize}
    \item The MHSA mechanism is the kernel of the transformer block, which captures dependencies between different positions in the input sequence through multiple attention heads \cite{vaswani2017attention}. Each attention head independently computes an attention score by performing linear transformations and scaled dot-product attention operations on three matrices, i.e., query $\mathbf{Q}$, key $\mathbf{K}$, and value $\mathbf{V}$, which can be mathematically expressed as 
    \begin{equation}
        \mathsf{Attention}(\mathbf{Q},\mathbf{K},\mathbf{V}) = \mathsf{softmax} \left( \frac{\mathbf{Q}\mathbf{K}^{\top}}{\sqrt{d_K}}\right) \mathbf{V},
    \end{equation}
    where $d_K$ is the dimension of $\mathbf{K}$. Then, all attention scores are concatenated and subjected to a linear transformation to yield the final attention output. The parallel computation nature of the MHSA mechanism allows the model to concurrently learn multiple representations across various subspaces, significantly enhancing the expressive capability of LLMs.
    \item The FFN consists of two linear transformations and a non-linear activation function (e.g., ReLU). It complements the MHSA mechanism by extracting deep features on each positional embedding, therefore aiding LLMs in capturing global dependencies. In particular, the FFN and the MHSA mechanism are often collectively referred to as a ``sublayer."
    \item The layer normalization typically follows the MHSA mechanism and FFN, which normalizes the input of each sublayer to stabilize and accelerate the training process. Residual connections mitigate the issue of vanishing gradients by summing the outputs of the MHSA mechanism and FFN with their inputs, hence preserving input features.
\end{itemize}

The synergistic operation of the three components enables the transformer block to capture global information and contextual dependencies within input sequences. By stacking multiple transformer blocks, LLMs can accurately understand contextual information and generate desired content that meets specific requirements. This attribute endows LLMs with substantial potential for supporting SemCom, offering exceptional capabilities in semantic understanding and content generation.
\subsection{Existing Applications of LLMs in SemCom}
The academic community has recently explored a variety of ways to integrate LLMs into SemCom, which can be categorized into two major applications, i.e., the direct use of LLMs for semantic coding and taking LLMs as auxiliary tools to facilitate semantic coding.
\subsubsection{\textbf{\textit{Direct use of LLMs for semantic coding}}} LLMs possess robust understanding and generation capabilities and have thus been utilized to perform semantic coding. For instance, Ref. \cite{wang2024uses} employed two pre-trained LLMs---bidirectional and auto-regressive transformers (BART) and GPT-2 as semantic codecs, in which a rate adaptation module was developed to align with the rate requirements of different channel codecs. Following it, Ref. \cite{chang2024gensc} adopted BART to perform bidirectional semantic coding. By capturing the correlation between consecutive tokens, BART enabled the receiver to recover semantically similar tokens, even when some tokens were corrupted or lost during transmission. In contrast, Ref. \cite{wang2024large} utilized the tokenizer training of LLMs to perform semantic encoding and leveraged unsupervised pre-training of LLMs to establish a KB, providing the receiver with prior knowledge for semantic decoding. In addition, several recent studies have utilized multimodal vision language models (VLMs), particularly bootstrapping language-image pre-training (BLIP) \cite{wei2024language,nam2024language,zhao2024lamosc,du2024generative} and contrastive language-image pre-training (CLIP) \cite{wang2024trustworthy} to perform efficient modality transformations between images and texts. By converting input images into concise textual prompts for transmission, the communication overhead can be significantly reduced. Moreover, an edge-device collaborative SemCom framework was proposed in Ref. \cite{ren2024generative}, which leveraged VLMs to generate textual prompts through visual captioning or question-answering, facilitating ultra-low-rate communications. Besides, Ref. \cite{cao2024multimodal} presented a multimodal LLM-based SemCom system, namely MLLM-PSC. It took text semantics as the key medium, enabling unified semantic conversion of diverse modalities without requiring additional KB alignment. 
\subsubsection{\textbf{\textit{Auxiliary tools for enhancing semantic coding}}} The extensive world knowledge of LLMs can serve as prior information to facilitate semantic coding \cite{zhao2024enhancing,jiang2024largeLLM}. For instance, an early study \cite{guo2024semantic} utilized BERT to perform importance-aware semantic understanding at the transmitter, and leveraged ChatGPT to conduct error calibration before semantic decoding. Ref. \cite{jiang2024visual} constructed a cross-modal KB using BLIP and stable diffusion, which extracted accurate text descriptions from input images to facilitate semantic coding. Moreover, a pre-trained large speech model---WavLM was employed to construct a semantic KB at the transmitter, enabling efficient semantic encoding and high-fidelity speech synthesis with low communication overhead \cite{zheng2024generative}. In addition to these, several pioneering studies \cite{jiang2024large0,yang2024rethinking,xie2024towards,jiang2024semanticFM,zhang2024addressing} have proposed multimodal LLM-enabled SemCom frameworks, where multimodal LLMs (e.g., GPT-4) served as semantic KBs for task decomposition, semantic representation, knowledge distillation, content calibration, transmission optimization, and out-of-distribution generalization. These frameworks make substantial contributions to standardized semantic encoding and personalized semantic decoding. In particular, by incorporating conditional GANs for channel estimation \cite{jiang2024large0} and RL for adaptive semantic offloading \cite{yang2024rethinking}, the reliability and scalability of SemCom systems were further enhanced. Finally, LLMs also demonstrated remarkable advantages in practical SemCom scenarios, such as satellite communications \cite{jiang2024semanticXXX}, underwater communications \cite{chen2024semantic}, and edge IoT networks \cite{kalita2024large}, where LLMs were utilized to perform semantic feature extraction, semantic importance assessment, and radio resource management, respectively. 
\begin{table*}[t]
\centering
\caption{An overview of four types of GAI-enabled SemCom systems.}
\label{tab:overview}
\begin{tabular}{p{1cm}|p{3.3cm}|p{3.5cm}|p{6cm}}
\toprule
\hline
{\textbf{Model}} & {\textbf{Structure}}   & {\textbf{Functions in SemCom}}  & {\textbf{References (classification by modality)}} \\ \hline
VAE   & \begin{tabular}[c]{@{}l@{}} Probabilistic encoder\\ Probabilistic decoder\end{tabular}  
      & Semantic coding/JSCC                               
      & \begin{tabular}[c]{@{}l@{}}Image:\cite{erdemir2022privacy, bo2024joint, alawad2022innovative, seon2024deep, zhang2024improving, zhang2024learned, li2024end, xie2024robust, hu2023robust, talli2024effective, si2024post, chen2024hierarchical}\\ Text:\cite{xi2024variational}\\ Audio: \cite{yao2023variational}\\ General: \cite{saidutta2020vae, ma2023task, nemati2023vq, choi2024semantics}\end{tabular}\\ \hline  
GAN   & \begin{tabular}[c]{@{}l@{}}Generator\\ Discriminator\end{tabular} 
      & \begin{tabular}[c]{@{}l@{}}Semantic coding/JSCC\\ Signal distortion suppression\end{tabular}          
      & \begin{tabular}[c]{@{}l@{}}Image: \cite{huang2022toward,han2023generative, huang2021deep, tang2024evolving, zhang2022deep, he2022robust, wang2022perceptual, xin2024deep,tan2024rate, erdemir2023generative, yu2024two, dong2022semantic, lokumarambage2023wireless, miao2024semantic, zhou2024moc, fu2023vector}\\ Text: \cite{mao2024gan} \end{tabular}\\ \hline
DM    & \begin{tabular}[c]{@{}l@{}}Forward diffusion process\\ Reverse denoising process\end{tabular}         
      & \begin{tabular}[c]{@{}l@{}}Semantic coding/JSCC\\ Channel modeling \\ Channel equalization\end{tabular}              
      & \begin{tabular}[c]{@{}l@{}}Image: \cite{jiang2024diffsc,grassucci2023generative, grassucci2024rethinking, pignata2024lightweight, zhang2024sc, fan2024semantic, wang2024fast, pezone2024semantic, fu2024multimodal, mingkai2024task, jiang2024adaptive, yang2024sg2sc, zhang2024diffusion, guo2024diffusion,chen2024commin,jiang2024large, pei2024latent, xu2023latent, wu2024cddm, li2024goal, zeng2024dmce, duan2024dm, yang2024rate, xie2024hybrid, ren2024diffusion, du2023ai, zheng2024energy, liu2024optimizing, du2023yolo, xu2024semantic,yuan2024generative} \\ Audio: \cite{qiang2024minimally,grassucci2024diffusion}\\ Multimodal: \cite{qiao2024latency, chen2024cross}    \end{tabular} \\ \hline 
LLM   & Transformer block                                                    
      & \begin{tabular}[c]{@{}l@{}}Semantic coding/JSCC\\ Auxiliary tool (e.g., KBs)\end{tabular}                        
      & \begin{tabular}[c]{@{}l@{}}Image: \cite{wang2024trustworthy, guo2024semantic, chen2024semantic, jiang2024largeLLM, zhao2024lamosc, zhang2024addressing, jiang2024visual, nam2024language, ren2024generative, jiang2024semanticXXX,du2024generative, wei2024language} \\ Text: \cite{wang2024large, wang2024uses, chang2024gensc, zhao2024enhancing} \\ Audio: \cite{zheng2024generative}\\ Multimodal: \cite{jiang2024large0, yang2024rethinking, kalita2024large, cao2024multimodal, jiang2024semanticFM, xie2024towards}\end{tabular} \\ 
     \toprule
\end{tabular}
\vspace{-6pt}
\end{table*}

\subsection{Challenges and Opportunities}
Thus far, we have systematically introduced four types of SemCom systems based on different GAI models, including VAEs, GANs, DMs, and LLMs. Table \ref{tab:overview} summarizes the key characteristics of these models alongside the relevant literature. Among these, there is an increasing belief in the research community that LLMs are the most promising technology for advancing intelligent SemCom. However, despite preliminary progress, the application of LLMs to SemCom still faces four critical challenges.
\begin{itemize}
    \item \textit{\textbf{Large modality gap}:} SemCom often involves multimodal data---such as images, audio, and videos---that extend beyond plain text. Each modality possesses unique data patterns, resulting in significant modality gaps that hinder LLMs from processing them efficiently. Although some studies have attempted to incorporate multimodal LLMs (e.g., GPT-4), their supported modalities remain limited, mainly to images and audio. It is still challenging to directly apply them to process higher-dimensional data, such as videos and 3D point clouds.
    \item \textit{\textbf{High adaptation cost}:} Current LLMs are primarily tailored for natural language tasks and do not account for the specific characteristics of SemCom, such as channel fading, interference, and noise. Applying existing LLMs directly to SemCom may lead to significant domain-specific biases. While fine-tuning LLMs can bridge the domain gap, adapting all model parameters incurs excessive communication and computation costs, which limit their practical deployment in resource-constrained wireless networks.
    \item \textit{\textbf{Limited generative capability}:} In the receiver design, most previous studies continue to follow the traditional mindset of information recovery, often leveraging specialized AI models (e.g., DMs) rather than establishing a new paradigm from a generative perspective. Such approaches fail to capitalize on the full generative potential of LLMs. Consequently, the limited exploration of the generative aspect restricts the system's ability to generalize to complex scenarios with diverse task requirements.
    \item \textit{\textbf{Inconsistent design philosophy}:} Prior works mainly employ LLMs to perform semantic encoding or construct semantic KBs, typically achieving incremental performance improvements. However, these approaches lack a unified guiding framework, resulting in fragmented methodologies that hinder the integration of research insights and limit the scalability of future SemCom systems.
\end{itemize}

To address these challenges, we propose a new LLM-native generative SemCom architecture in the next subsection. The key novelties are summarized as follows.
\begin{itemize}
    \item \textit{\textbf{Mitigating the modality gap}:}  We introduce a perception encoder that projects multimodal input data into the same feature space as language tokens. This enables LLMs to efficiently understand and process diverse forms of data.
    \item \textit{\textbf{Reducing the adaptation cost}:} We introduce a lightweight fine-tuning module that functions as a ``thin wrapper" for LLMs. By freezing the full parameters of the pre-trained LLM and training only the fine-tuning module, the general knowledge of the LLM is well preserved while the SemCom domain knowledge can be efficiently acquired. 
    \item \textit{\textbf{Enhancing the generative capability}:} We integrate both LLMs and specialized AI models in the receiver design. LLMs are employed for content generation while specialized AI models refine the outputs to eliminate uncertainty and hallucination. This hybrid design leverages the strengths of both model types to unleash the generative potential at the receiver. 
    \item \textit{\textbf{Unifying the design philosophy}:} We develop a universal AI agent framework for both the transmitter and receiver, in which LLMs serve as the core to perform either information understanding or content generation. This framework ensures deep integration of LLMs within SemCom systems, moving beyond their current role as auxiliary components.
\end{itemize}

Before presenting the detailed designs, we emphasize that the proposed generative SemCom architecture represents a paradigm shift in communication, transitioning from traditional ``information recovery" to ``information regeneration." By exploiting the generative aspect of LLMs at the receiver, our architecture establishes a coherent and universal design methodology that supports a “one architecture for all scenarios” vision,  fostering the sustainable development of generative SemCom systems.
\subsection{LLM-native Generative SemCom Architecture}
\begin{figure*}[t]
\centering
\includegraphics[width=0.9\linewidth]{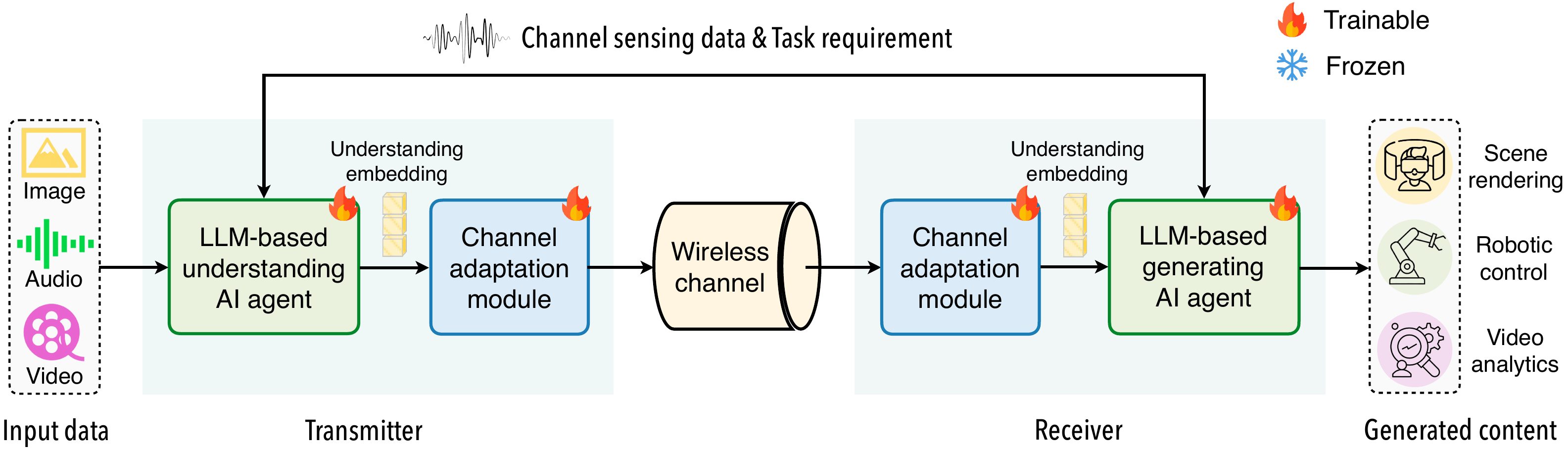}
\caption{LLM-native generative SemCom architecture. The LLM-based AI agent at the transmitter performs information understanding to extract a compact understanding embedding in the form of discrete tokens. Once receiving the (noisy) understanding embedding, the LLM-based AI agent at the receiver directly generates the desired content.} \label{fig:LLM_architecture}
\vspace{-8pt}
\end{figure*}
\begin{figure*}[t]
\centering
\includegraphics[width=0.71\linewidth]{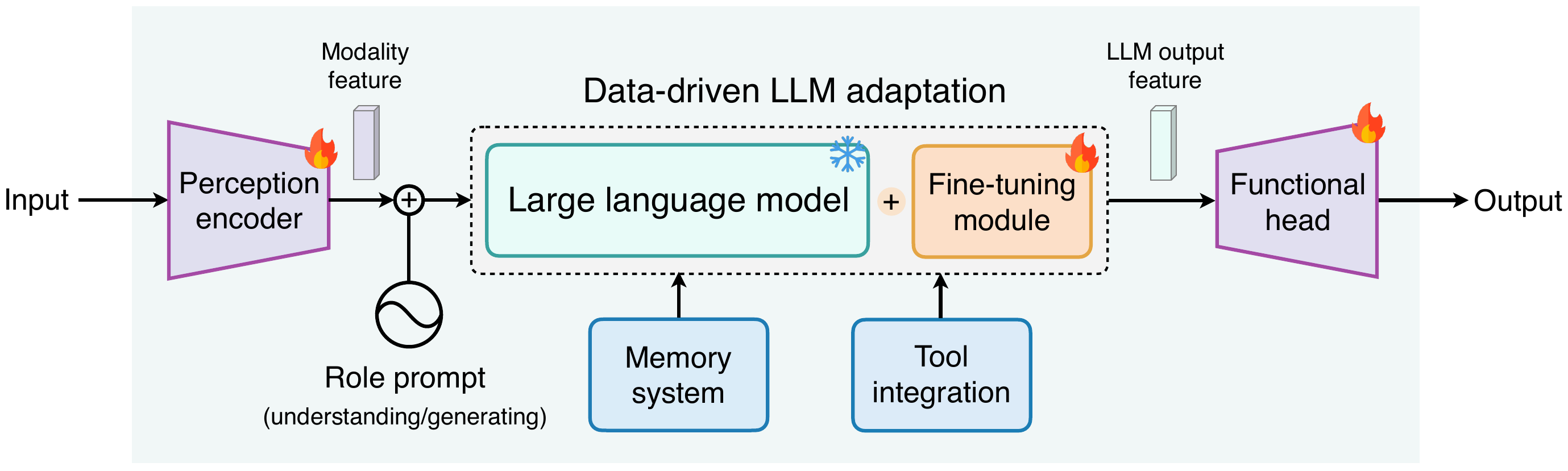}
\caption{The framework of LLM-based AI agents. The understanding and generating AI agents are delineated by a role prompt. For the understanding AI agent, its input includes the original data, channel sensing data, and task requirement, while its output is the understanding embedding. For the generating AI agent, its input includes the received (noisy) understanding embedding, channel sensing data, and task requirement, while its output is the generated content. The tool integrations and functional heads of the two AI agents are configured as needed.} \label{fig:AI_agent}
\vspace{-8pt}
\end{figure*}
\subsubsection{Architecture Overview}
As illustrated in Fig. \ref{fig:LLM_architecture}, the LLM-native generative SemCom system comprises four building blocks, with the transmitter and receiver each equipped with an LLM-based AI agent and a channel adaptation module. Drawing an analogy to human-to-human conversation, the two LLM-based AI agents serve as the ``brains" of interlocutors while the two channel adaptation modules act as their ``mouths and ears." The operational mechanism of the system is described as follows.
\begin{itemize}
    \item Initially, the LLM-based AI agent at the transmitter comprehends the input data in conjunction with the channel sensing data and task requirement, yielding a compact semantic representation, namely \textit{understanding embedding}. Notably, this embedding encapsulates the information essential for accomplishing the receiver's task, which is seen as the ``prompt" for the AI agent at the receiver.
    \item Next, the channel adaptation module at the transmitter transforms the understanding embedding into a signal suitable for wireless transmission.
    \item After receiving the noisy signal, the channel adaptation module at the receiver converts it back into the understanding embedding.
    \item Finally, the LLM-based AI agent at the receiver directly generates task-oriented content by leveraging the received understanding embedding.
\end{itemize}

In the following, we elaborate on the detailed design of each building block.
\subsubsection{LLM-based AI Agent} There is a growing consensus that LLM-based AI agents are the next frontier in the field of GAI \cite{durante2024agent}. Considering its robust capabilities in information analytics and human-like responses, we deploy two LLM-based AI agents at the transmitter and receiver for information understanding and content generation, which are referred to as the {\it understanding AI agent} and {\it generating AI agent}, respectively. As depicted in Fig. \ref{fig:AI_agent}, both AI agents adopt the same backbone structure consisting of five components, including a perception encoder, a data-driven LLM adaptation, a memory system, a tool integration, and a functional head. In particular, the two AI agents are delineated by an embedded ``role prompt" and can achieve multi-agent collaboration through end-to-end training. The function of each component is described below.

\textit{~~a) \textbf{Perception encoder}} is responsible for sensing communication environments and preprocessing input data to enable the efficient utilization by LLMs. Its input encompasses multimodal data (e.g., images, audio, videos, or their understanding embeddings), channel sensing data (e.g., CSI and environmental point cloud data), and task requirements (e.g., scene rendering, robotic control, and video analytics). Since most LLMs only accept plain text as input, it is necessary to project all data into a unified feature space as language tokens and represent them as token embeddings for LLM utilization. A simple way to achieve this goal is to use existing well-designed encoders tailored for specific modalities, such as the vision transformer (ViT) for images and the C-Former for audio. Several recent studies have also proposed unified encoders to handle multiple modalities, such as the ImageBind \cite{girdhar2023imagebind} and SBER-MoVQGAN5 \cite{wang2024emu3}, achieving ``one encoder for all modalities." Instead of handcrafting the perception encoder from scratch, reusing these ``out-of-the-box" models significantly reduces engineering costs. In particular, a linear mapping layer could be added subsequent to each modality encoder to align the dimensionality of the modality feature with that of the LLM's token space.

\textit{~~b) \textbf{Data-driven LLM adaptation}} performs information understanding at the transmitter and content generation at the receiver, respectively. It is composed of a pretrained LLM (e.g., ChatGPT-3.5 and Llama 3.2) and a fine-tuning module. The LLM excels in understanding data and generating new content, while the fine-tuning module is optimized to achieve LLM adaptation to SemCom. Several popular parameter-efficient fine-tuning methods can be employed, such as low-rank adaptation, prompt tuning, prefix tuning, and adapter tuning \cite{han2024parameter}. By freezing the  LLM's parameters and learning a few external parameters through a data-driven pipeline, the fine-tuning module enables the LLM to acquire SemCom domain knowledge with minimal cost. Notably, the objective of data-driven LLM adaptation is controlled by a role prompt. For instance, the role prompt at the transmitter could be ``[understanding prompt] = please act as a (domain) expert, understand the input information and provide an accurate description.” Correspondingly, the role prompt at the receiver is  ``[generating prompt] = please generate content based on the input information to meet the task requirement.” It is worth mentioning that for complex tasks, the LLM can leverage its chain-of-thought to decompose tasks into hierarchical levels for iterative processing, thereby improving system efficiency.

\textit{~~c) \textbf{Memory system}} is akin to the “hippocampus” in human brains, which accumulates knowledge during training to assist the LLM in fast information understanding and content generation. Depending on the timeliness of stored knowledge, the memory system is categorized into short-term memory and long-term memory. Short-term memory contains contextual information from instant communications to ensure content coherence, which is generally implemented through caching and recurrent neural networks. In contrast, long-term memory distills general knowledge from historical data and organizes it in a structured manner, such as knowledge graphs and vector databases. Short-term memory and long-term memory are mutually complementary to enhance the efficiency of AI agents. For example, when encountering identical tasks during communications, the LLMs at the transceiver can retrieve relevant knowledge fragments from the associated memory systems to achieve light-inference or even free-inference. Furthermore, through efficient knowledge refinement, the memory system may capture the preferences and behavior patterns of the transceiver, which in turn facilitates the self-evolution of AI agents.
    
\textit{~~d) \textbf{Tool integration}} serves as the ``extended arm" of the LLM. It integrates various professional tools and application programming interfaces (APIs)---such as search engines and data access interfaces---to support the LLM in information understanding and content generation in specialized domains. For example, in video transmission scenarios, the ``video intelligence API" can be invoked to extract key scenes, events, and target objects from the original video, thereby facilitating the LLM to produce a more accurate video summary. Meanwhile, the LLM at the receiver can utilize the ``RunwayML" tool to generate video content that aligns with the video summary. These tools and APIs overcome the LLM's understanding biases and hallucination effects in specialized domains, significantly enhancing the effectiveness and reliability of the generated content. In particular, due to the distinct responsibilities of the understanding and generating AI agents, they are configured with different tools and APIs in general.

\textit{~~e) \textbf{Functional head}} aims to process the LLM's output feature to achieve the ultimate goal of the AI agent. On one hand, the understanding AI agent needs to output in-depth understanding information about the input data. A feasible approach is to utilize the classical language modeling head. It predicts tokens in an autoregressive manner and eventually combines all tokens into the understanding embedding. The result is a compact language description and is much smaller than the original data. In particular, this ``next-token prediction" method offers a unified representation for any-to-any tasks with high compatibility \cite{wang2024emu3}. On the other hand, the generating AI agent has to render the LLM's output feature into the task-oriented content. Consequently, specialized generative models can be customized according to task requirements, such as stable diffusion for image reconstruction, AudioLDM for speech synthesis, and Zeroscope for video prediction \cite{zhang2024mm}. These models are able to generate desired content through a single step, albeit with some additional storage overhead.

\subsubsection{Channel Adaptation Module} To achieve reliable transmission of the understanding embedding, we deploy two channel adaptation modules at the transmitter and receiver, respectively. Specifically, the channel adaptation module at the transmitter converts the understanding embedding into a signal that matches the wireless channel, while the counterpart channel adaptation module at the receiver transforms the received noisy signal back into the understanding embedding. Essentially, they involve two mapping processes that can be realized using classical discriminative models, such as convolutional neural networks. Nevertheless, these models are short of adaptability and generalization, making it challenging to cope with the dynamics of wireless channels (e.g., channel fading) and the diversity of communication models (e.g., MIMO). To address this issue, an alternative approach is to employ GAI models, particularly GANs and DMs. Given their strong self-learning abilities, GAI-enabled channel adaptation modules can well adapt to dynamic communication environments via continuous learning. Moreover, by leveraging the self-supervised learning mechanism, they can perform information importance assessment, wireless resource allocation, and network parameter optimization to further enhance communication efficiency and task success rates \cite{liang2024generative}.

In summary, the two LLM-based AI agents perform data understanding and content generation, respectively. Their synergistic collaboration enables efficient data transformation between high-dimensional and low-dimensional spaces, significantly reducing the communication overhead and enhancing the generalization capability of the SemCom system. Meanwhile, the two channel adaptation modules act as ``bridges" between the two AI agents, ensuring reliable information transmission over wireless channels. This seamless integration harnesses the full power of LLMs, laying a solid foundation for building efficient generative SemCom systems in future wireless networks.

\section{A Case Study of LLM-native Generative SemCom}
Building upon the system designs, this section will provide a case study to validate the effectiveness of the proposed generative SemCom system. 
\subsection{Experimental Settings}
\subsubsection{Task Requirements} We consider a point-to-point communication system designed for a video retrieval task. In the system, the transmitter possesses a large video, while the receiver aims to identify and retrieve specific clips of the video that contain particular objects of interest, such as people and vehicles. Such tasks are prevalent in domains of security surveillance, traffic monitoring, and remote sensing object tracking. Given the massive data volume of the original video and the receiver's interest only in specific clips, SemCom is well-suited for these scenarios, promising to significantly enhance the system efficiency.
\subsubsection{System Configurations}
The default system configurations are as follows. The transmitter employs the YOLOv8-DeepSORT-Object-Tracking model \cite{faisal2023yolov8} as the perception encoder, which performs object tracking on all objects in the original video and extracts associated keyframes. Additionally, the transmitter is equipped with a multimodal LLM---InternVL-1.5 \cite{chen2023internvl}. It processes each keyframe to generate a sequence of descriptive entries, forming the understanding embedding. To guide InternVL-1.5 in generating standardized entries, we design the following understanding prompt.

\begin{mdframed}[linewidth=0.8pt,linecolor=black]
{\noindent {\bf[Understanding Prompt]:} You are a video attribute extraction expert. Your task is to analyze a given video frame and accurately extract the attributes related to a person's appearance. Please provide the output as a concise list of keywords following this format as output. Ensure the output is a single line and does not span multiple lines: {\it [gender], [shirt color], [shirt length], [pants color], [pants length], [shoe color]}.}
\end{mdframed}

At the receiver, another multimodal LLM---Qwen \cite{qwen2}, is deployed, which generates a complete description of the keyframe based on the received entries. To standardize the output from Qwen, we design the following generating prompt.
\begin{mdframed}[linewidth=0.8pt,linecolor=black]
\noindent {\bf [Generating Prompt]:}
You are a semantic expansion assistant. Given an input in the format: {\it [time], [location], [gender], [shirt color], [shirt length], [pants color], [pants length], [shoe color]}, your task is to expand it into a complete sentence using the following structure:
{\it ``This is a [gender] person wearing a [shirt color] [shirt length] shirt, [pants color] [pants length] pants, and [shoe color] shoes. [He/She] appeared at [location] at [time]."}
\end{mdframed}
\noindent In addition, the receiver employs all-MiniLM-L6-v2 model \cite{allMiniML6v2} as the functional head to generate retrieval results, i.e., to determine whether the target object is present in the keyframe based on the Qwen’s output. Besides, the transmitter and receiver are connected through an AWGN channel. Their channel adaptation modules are implemented using two linear mapping layers.
\subsubsection{Baseline Schemes} To demonstrate the superiority of the proposed LLM-native generative SemCom system, four baseline schemes are adopted for performance comparison, including
\begin{itemize}
    \item MPEG + 1/2 LDPC + 16QAM, which employs the moving picture experts group (MPEG) compression method for source coding, while utilizing the rate-1/2 low-density parity-check (LDPC) code and 16-quadrature amplitude modulation (QAM) for channel coding and modulation, respectively. 
    \item VAE-enabled scheme, with a pre-trained VAE \cite{ liu2020unsupervised} as the backbone for JSCC.
    \item GAN-enabled scheme, with a pre-trained GAN \cite{ wang2022realesrgan} as the backbone for JSCC.
    \item DM-enabled scheme, with a pre-trained DM \cite{lin2023diffbir} as the backbone for JSCC.
\end{itemize}

For each baseline scheme, when the receiver recovers the video frames, a widely-employed object retrieval algorithm \cite{lin2019improving} is applied for decision-making. In particular, the first baseline is referred to as the traditional communication scheme, while our scheme is named as LLM-native generative scheme for brevity.
\subsubsection{Performance Metrics} 
In this case study, we use three metrics to evaluate the system performance, including (1) Retrieval accuracy, which refers to the proportion of keyframes that contain the target object among all retrieved keyframes; (2) Robustness, measured by the standard deviation of the retrieval accuracy across varying signal-to-noise ratios (SNRs), reflecting the system's sensitivity to channel noise; (3) Communication overhead, quantified by the total number of bits being transmitted.
\begin{figure}[t]
\centering
\includegraphics[width=0.98\linewidth]{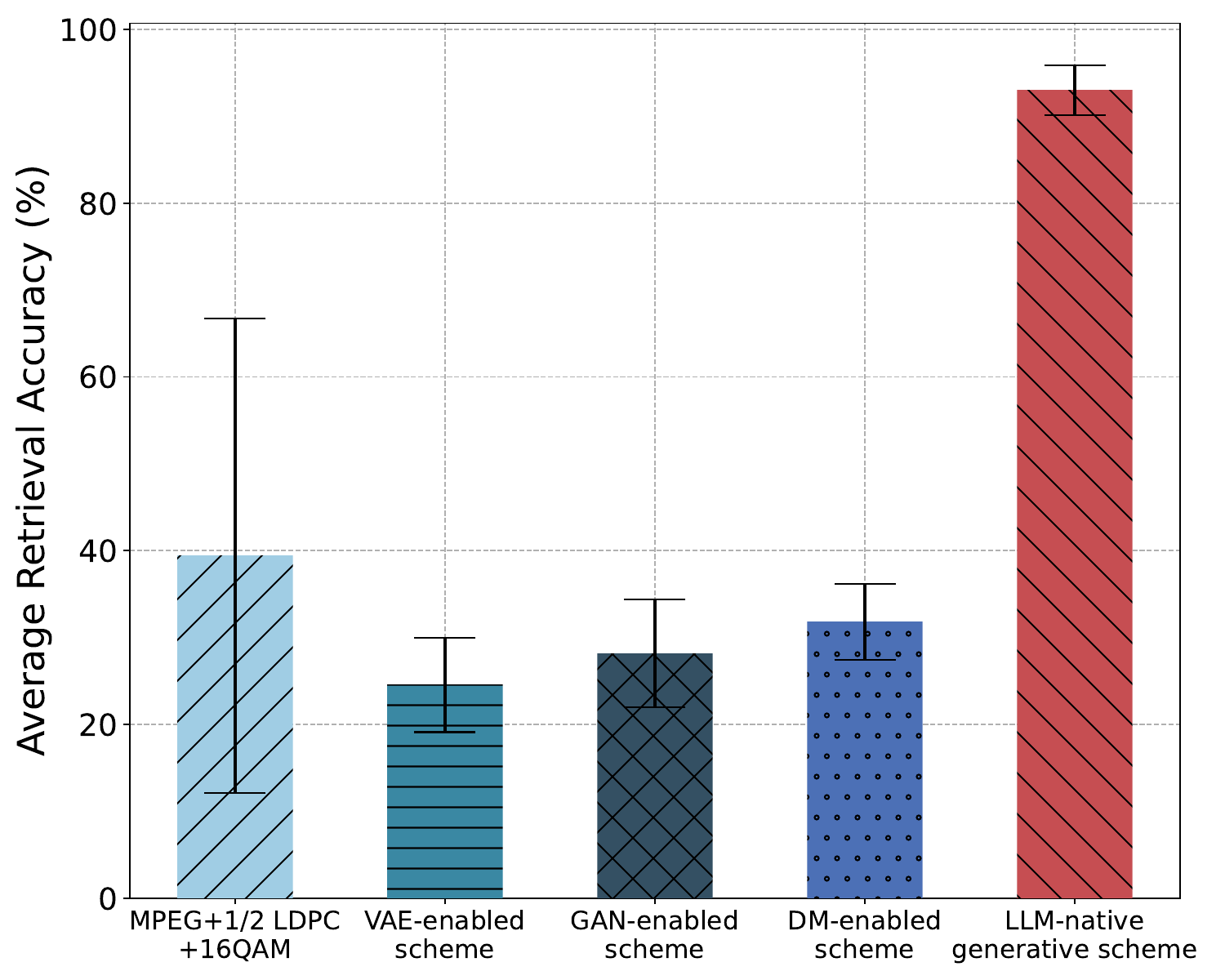}
\caption{Comparison of the retrieval accuracy across the five schemes. The height of each bar represents the average retrieval accuracy, while the error bar indicates the standard deviation of the retrieval accuracy.} \label{fig:experiment}
\vspace{-10pt}
\end{figure}
\subsection{Results and Analysis}
We conduct experiments at different SNR regimes ranging from 3 dB to 30 dB, incrementing by 3 dB each time. Fig. \ref{fig:experiment} shows the performance comparison of the five schemes, where the height of each bar represents the average retrieval accuracy and the error bar indicates the standard deviation of the retrieval accuracy. A longer error bar indicates higher variability in retrieval accuracy. As observed from Fig. \ref{fig:experiment}, we obtain the following key insights.
\subsubsection{Retrieval Accuracy}
The LLM-native generative scheme achieves an average retrieval accuracy of 93.03\%, significantly outperforming all baseline schemes. Compared to the traditional communication scheme (i.e., MPEG + 1/2 LDPC + 16QAM), which achieves an average retrieval accuracy of 39.39\%, the LLM-native generative scheme yields an approximate gain of 53\%. This improvement is attributed to the LLM's strong generalization capability. More precisely, compared to the small-scale object retrieval algorithm used in the baseline schemes, LLMs can identify more accurate attributes of the target object in the keyframes. Furthermore, the standardized outputs of the two LLMs enhance the retrieval accuracy as well. Notably, the VAE-enabled scheme has the lowest average retrieval accuracy. This is because VAE's probabilistic generation method leads to low-quality generated frames, which in turn degrades retrieval performance.
\subsubsection{Robustness} The LLM-native generative scheme has the smallest error bar, showcasing superior robustness against channel noise. This result benefits from the LLM's extensive knowledge, which effectively compensates for the noise effect. In contrast, the traditional communication scheme exhibits the largest error bar, implying the weakest noise resistance. The reason is that the traditional communication scheme suffers from a ``cliff effect," wherein the receiver fails to recover keyframes when the SNR is below a certain threshold (9 dB in our experiment). Conversely, when the SNR exceeds the threshold, it restores keyframes very well, resulting in a bimodal performance distribution. Besides, the DM-enabled scheme demonstrates higher robustness than the VAE-enabled and GAN-enabled schemes due to its stable training process.
\begin{table}[t]
\centering
\caption{Communication overhead of the five schemes.}
\label{tab:communication}
\begin{tabular}{l|l}
\toprule
\hline
Scheme                        & Communication overhead \\ \hline
MPEG+1/2 LDPC+16QAM           & 219 Mbits              \\ \hline
VAE-enabled scheme            & 13 Mbits               \\ \hline
GAN-enabled scheme            & 13 Mbits               \\ \hline
DM-enabled scheme             & 13 Mbits               \\ \hline
LLM-native  generative scheme & {\bf 36 Kbits}         \\ 
\toprule
\end{tabular}
\end{table}
\begin{figure*}[t]
\centering
\includegraphics[width=0.9\linewidth]{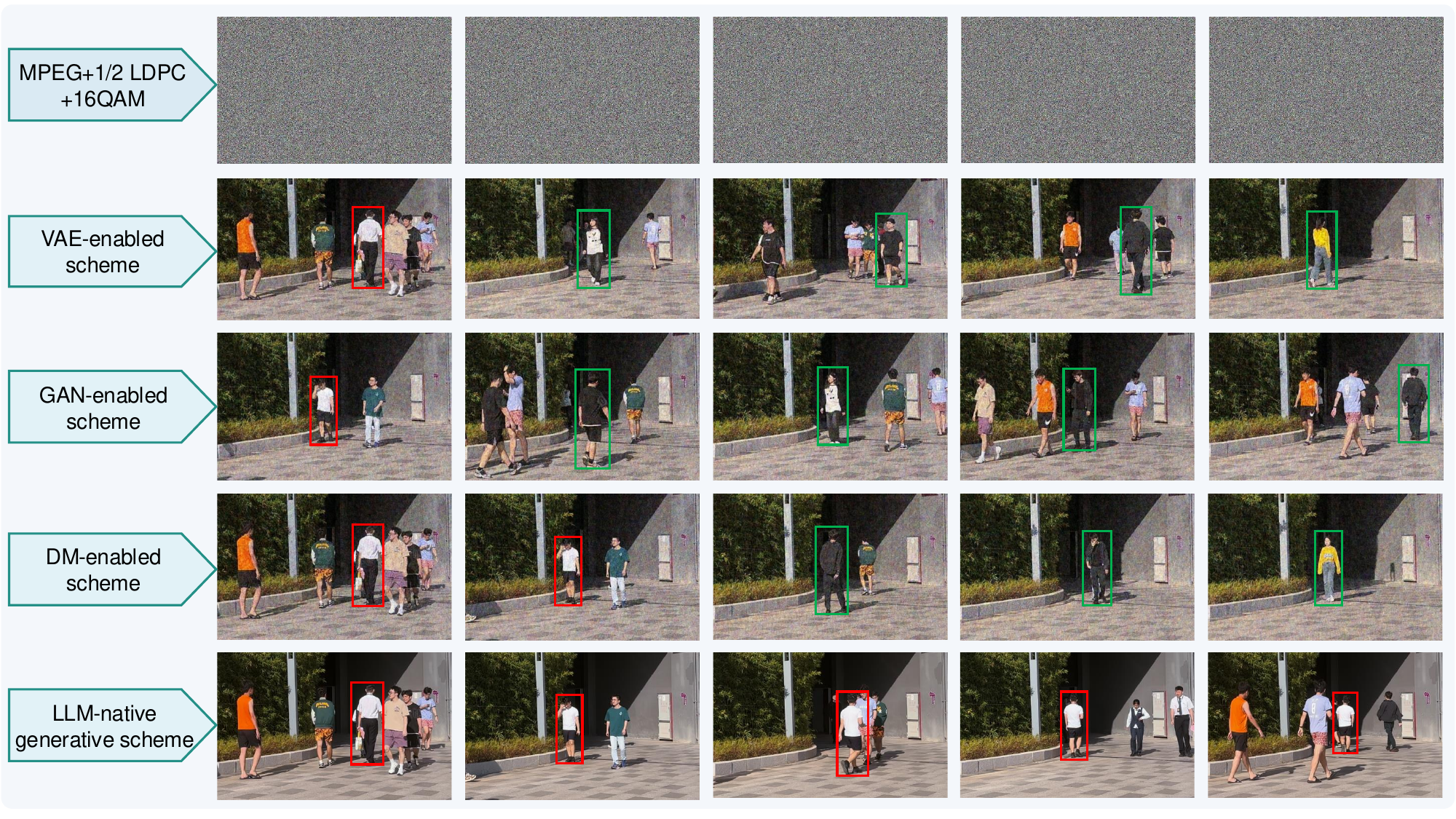}
\caption{Visual comparison of the five schemes when SNR = 9 dB. The \textcolor{red}{red boxes} show correct retrieval results while the \textcolor{darkgreen}{green boxes} indicate wrong retrieval results.}
\label{fig:visualization}
\end{figure*}
\subsubsection{Communication Overhead}
Table \ref{tab:communication} provides the communication overhead for all schemes. The LLM-native generative scheme demonstrates a communication overhead that is orders of magnitude lower, reducing the transmission volume by 99.98\% compared to the traditional communication scheme. This result is quite intuitive, as the LLM-native generative scheme transmits only concise entries rather than raw videos or high-dimensional semantic features.
\subsubsection{Visualization Results}
Fig. \ref{fig:visualization} presents the retrieval samples of each scheme, where the SNR is set to 9 dB. The task requirement is---identifying the man wearing a white long-sleeved shirt, black pants, and black shoes in the video. It is observed that the keyframes retrieved by the LLM-native generative scheme all contain the target object (in red), whereas other schemes make retrieval errors (in green). In addition, the traditional communication scheme fails to recover keyframes at the receiver due to the low SNR.

The above experimental results validate that the LLM-native generative scheme outperforms other baseline schemes in terms of retrieval accuracy, robustness, and communication overhead. These findings underscore the potential of LLM-native generative SemCom for diverse applications in future wireless networks.
\section{Promising Applications of Generative SemCom}
In this section, we summarize four practical applications for generative SemCom, including industrial IoT (IIoT), vehicle-to-everything (V2X), metaverse, and low-altitude economy.
\subsection{Industrial Internet of Things}
The IIoT enables flexible resource configurations and sustainable manufacturing production through the collaboration and interoperability of various industrial elements, including humans, sensors, and machines \cite{sisinni2018industrial}. This integration necessitates the industrial elements to exchange a large amount of data, such as production indicators, environmental information, and machinery operating statuses. However, existing industrial communication technologies cannot efficiently integrate manufacturing intentions with data communications, leading to excessive signaling overhead, low collaboration efficiency, and severe security risk. By employing the proposed generative SemCom architecture, the transmitter on the production line can utilize the LLM-based AI agent to conduct an in-depth understanding of the input data and provide compact understanding information aligned with the manufacturing objective. According to the understanding information, the receiver on the production line can directly generate appropriate operational instructions, such as adjusting production schedules and performing machine maintenance, without the need to recover the original data. This convergence of communication and computation significantly streamlines the manufacturing process and enhances production efficiency, thus advancing the intelligent transformation of industrial manufacturing.
\subsection{Vehicle-to-Everything}
The V2X aims to enhance driving comfort, safety, and traffic efficiency through the interconnection among vehicles, humans, and infrastructure. The implementation of V2X in practical systems requires ultra-low communication latency, real-time data processing, and robust privacy protections. Generative SemCom offers a viable solution to meet these stringent requirements \cite{lu2024generativeV2X}. By utilizing local LLM-based AI agents, vehicles can process multimodal sensor data---including camera feeds and LiDAR signals---for real-time road condition analytics, such as obstacle detection and traffic light recognition. The analytical results enable the generation of control commands for steering, acceleration, and braking, hence supporting autonomous driving. Meanwhile, AI agents can engage with users to enhance their driving experiences. Furthermore, generative SemCom supports instantaneous communications between vehicles and infrastructures. Vehicles can broadcast their surrounding information to assist nearby vehicles in route optimization and roadside infrastructures in traffic control and scheduling, which is particularly valuable in latency-sensitive scenarios such as highways and congested intersections. Notably, since raw data is not communicated, user privacy and security are also strengthened.
\subsection{Metaverse}
The metaverse signifies a seamless integration of the physical and digital worlds,  aspiring to create a virtual space that fosters immersive interaction and collaboration among users. Realizing this vision necessitates advanced communication technologies capable of delivering millisecond latency and secure information exchange \cite{lin2023unified}. Traditional communication systems struggle to meet these demands, particularly when dealing with multimodal interactions and dynamic environmental updates. Generative SemCom provides a potential solution by deploying LLM-based AI agents in metaverse devices \cite{huang2024privacy}. Since only semantic information is exchanged between devices, the communication efficiency and security will be significantly enhanced. For instance, during human-avatar interactions, the transmitter's AI agent can analyze user actions and abstract them into semantic descriptions for transmission, while the receiver's AI agent generates smooth and contextually consistent actions in real time based on the received descriptions. Similarly, for virtual environment creation, users can exchange concise scene descriptions rather than detailed geometric data, thus enriching the collaborative experience. Generative SemCom can also support other metaverse applications, such as intelligent interactions with non-player characters and personalized digital content generation by mining user preferences \cite{kurai2024magicitem}.
\subsection{Low-altitude Economy}
The low-altitude economy capitalizes on the airspace below 1,000 meters for economic activities, particularly through the use of flying devices such as unmanned aerial vehicles (UAVs). It is projected that tens of thousands of flying devices will populate urban airspaces, necessitating the provision of ultra-low latency, extensive connectivity, and anti-interference communication services to ensure efficient and orderly management. Generative SemCom is poised to meet these requirements by deploying LLM-based AI agents on flying devices, which transform massive traffic data into concise semantic information, thus alleviating the total traffic load on low-altitude networks. For instance, in smart logistics, UAVs can send environmental information to a centralized platform after thorough analytics, aiding in real-time scheduling and decision-making. In video surveillance, UAVs can analyze the videos based on user requirements (e.g., searching for individuals) and transmit task-relevant information for immediate response. Additionally, in smart agriculture, the AI agents on UAVs can assess crop health and identify areas of concern by analyzing multispectral images. The resulting information facilitates ground equipment to generate precise instructions for interventions, such as pesticide application, thereby improving the timeliness of agricultural practices.
\section{Open Issues and Future Directions}
Generative SemCom remains in its nascent stages, with numerous technical and engineering challenges yet to be resolved. In this section, we outline three primary challenges associated with generative SemCom and discuss potential solutions to address these issues.
\subsection{Deployment of LLMs on Resource-Constrained Devices}
The deployment of LLMs on resource-constrained edge devices is the primary challenge for generative SemCom due to the substantial storage requirements, high energy consumption, and long computational latency. Firstly, the parameter count of existing LLMs ranges from billions to trillions (e.g., 1.23 billion for Llama 3.2-1B and 1.75 trillion for GPT-4), necessitating tens of GigaBytes to several TeraByte of storage space for on-device deployment. This requirement is excessively stringent for edge devices, which have only a few hundred GigaBytes of storage resources in general. Secondly, the training and inference processes of LLMs involve extensive floating-point operations, which are prohibitive for battery-powered edge devices in terms of energy consumption. Furthermore, the computational limitations of edge devices lead to significant latency (typically several seconds) when utilizing LLMs for information understanding and content generation, failing to meet the real-time requirements of latency-sensitive applications like holographic communication. To address these issues, future research should focus on the development of lightweight on-device LLMs. A potential solution is to design new model pruning techniques (e.g., LLM-pruner \cite{ma2023llm}) and knowledge distillation methods (e.g., MiniLLM \cite{gu2024minillm}) to reduce model size while maintaining model performance. Another solution is to collaborate the storage and computational resources at edge servers for distributed LLM deployment \cite{xue2024wdmoe}. Additionally, the customization of LLMs for specific scenarios could be investigated to achieve the balance between model size, computational efficiency, and system performance \cite{wu2024netllm}.
\subsection{Dynamic Evolution of AI Agents at Transceivers}
The collaboration between the understanding AI agent and generating AI agent is crucial for ensuring the performance of generative SemCom systems. However, the variability inherent in wireless networks---characterized by multimodal data, dynamic channels, and diverse tasks---poses significant challenges to this collaboration. On one hand, the uncertainty in channels and the complexity of tasks may introduce biases in the understanding AI agent, which compromises the accuracy of the content generated by the generating AI agent. On the other hand, the knowledge discrepancies between the two AI agents are likely to become exacerbated over time, adversely affecting the overall system performance. These challenges necessitate the dynamic evolution of the two AI agents. For one thing, it is advisable to incorporate popular continual learning approaches, such as experience replay and meta-learning \cite{wang2024comprehensive}. These methods enable the ongoing refinement of AI agents by learning patterns from newly acquired data, thus enhancing their adaptability to variations in communication environments and task requirements. For another thing, developing new knowledge-sharing protocols and bidirectional feedback mechanisms is promising to achieve efficient knowledge synergy between the transmitter and receiver~\cite{li2024cooperative}. For example, the transmitter might employ knowledge distillation techniques to selectively share crucial insights with the receiver, assisting its AI agent in generating more effective content. In turn, the receiver could provide feedback on the quality of content to the transmitter, which prompts its AI agent to optimize the understanding process, hence fostering a cycle of self-improvement within the system.
\subsection{Privacy and Security Concerns During Transmission}
The integration of LLMs into SemCom systems raises significant privacy and security concerns. These issues primarily stem from the propensity of LLMs to memorize data information during their training phases~\cite{wang2024reinforcement}. For instance, the understanding AI agent at the transmitter may unconsciously encode partial information of the training data into its output, i.e., the understanding embedding. As a result, unauthorized third parties could intercept the communication signals and leverage the same LLM to reconstruct training data or conduct membership inference attacks. Furthermore, the generative philosophy introduces potential risks that malicious entities could exploit. Specifically, these entities might manipulate the understanding embedding through poisoning attacks, leading to the generation of erroneous or misleading content at the receiver. Such vulnerabilities have severe implications for security-sensitive applications, such as virtual medical consultations and confidential business communications~\cite{jiang2024poster}. To address these issues, it is essential to develop advanced encryption technologies (e.g., quantum-resistant algorithms), and efficient privacy-preserving mechanisms (e.g., differential privacy), to prevent the understanding embedding from unauthorized access. Moreover, incorporating adversarial training methods during LLMs' training phases may enhance their resilience against poisoning attacks~\cite{yu2024robust}. Furthermore, physical layer security techniques---such as secure beamforming strategies and covert communication algorithms~\cite{xu2024covert}---can be employed to bolster the system's confidentiality. Besides, future research should focus on establishing a secure framework for multiuser generative SemCom, wherein blockchains can be weaponized for secure recording and robust verification of semantic transmission, safeguarding the system against potential threats \cite{lin2024blockchain}.
\section{Concluding Remarks}
This paper presented a comprehensive overview and research outlook on the integration of GAI with SemCom. We began by introducing three types of SemCom systems enabled by classical GAI models, including VAEs, GANs, and DMs. Then, we proposed a novel LLM-native generative SemCom system, which was characterized by two LLM-based AI agents at both the transmitter and receiver, signifying a pivotal shift in the communication mindset---from ``information recovery" to ``information regeneration." To substantiate this system design, we conducted an initial case study to demonstrate the benefits derived from LLM utilization. Furthermore, we identified four practical applications for generative SemCom, including IIoT, V2X, metaverse, and low-attitude economy, showcasing its real-world potential. Finally, three open issues along with their preliminary solutions were discussed. 

We believe that the integration of advanced GAI models, particularly LLMs, promises to redefine the communication paradigm from a generative perspective. This study hopefully serves as a valuable reference and provides insightful guidelines for further in-depth investigations of generative SemCom in the context of 6G.

\bibliographystyle{IEEEtran}
\bibliography{reference}

\end{document}